# Suction-based Soft Robotic Gripping of Rough and Irregular Parts

Sukho Song[1,2]*, Dirk-Michael Drotlef[1]*, Donghoon Son[1], Anastasia Koivikko[1,3], and Metin Sitti[1†]

[1]Physical Intelligence Department, Max Planck Institute for Intelligent Systems, 70569 Stuttgart, Germany
[2]Laboratory for Soft Bioelectronic Interfaces, École Polytechnique Fédérale de Lausanne, 1202 Geneva, Switzerland
[3]Faculty of Medicine and Health Technology, Tampere University, 33720 Tampere, Finland

[†]Correspondence to: sitti@is.mpg.de
*Equally contributing first authors

## Abstract

Recently, suction-based robotic systems with microscopic features or active suction components have been proposed to grip rough and irregular surfaces. However, sophisticated fabrication methods or complex control systems are required for such systems, and robust attachment to rough real-world surfaces still remains a grand challenge. Here, we propose a fully soft robotic gripper, where a flat elastic membrane is used to conform and contact parts or surfaces well, where an internal negative pressure exerted on the air-sealed membrane induces the suction-based gripping. 3D printing in combination with soft molding techniques enable the fabrication of the soft gripper. Robust attachment to complex 3D and rough surfaces is enabled by the surface-conformable soft flat membrane, which generates strong and robust suction at the contact interface. Such robust attachment to rough and irregular surfaces enables manipulation of a broad range of real-world objects, such as an egg, lime, and foiled package, without any physical damage. Compared to the conventional suction cup designs, the proposed suction gripper design shows a four-fold increase in gripping performance on rough surfaces. Furthermore, the structural and material simplicity of the proposed gripper architecture facilitates its system-level integration with other soft robotic peripherals, which can enable broader impact in diverse fields, such as digital manufacturing, robotic manipulation, and medical gripping applications.

## Summary

A fully soft suction gripper is proposed to achieve both high surface conformation and strong and robust gripping on rough and irregular surfaces.

## Introduction

Robust, yet reversible adhesion mechanisms on rough and irregular surfaces remain an ongoing scientific challenge for artificial smart adhesives, which can have broad applications in medical devices and adhesives (*1-3*), digital manufacturing (*4, 5*), robotic manipulation (*6-11*), and transfer printing (*12-17*). Biological systems have evolved to create various robust solutions to achieve strong and controllable adhesion properties on real-world surfaces, e.g., geckos using intermolecular forces for dry adhesion (*18-20*), ants, beetles, and tree frogs relying on the capillary force (*21-23*), and octopi, chameleons, and clingfishes adhering to a wide range of rough surfaces by suction (*24-26*). Various synthetic adhesive materials have been reported inspired by these biological attachment systems, such as gecko-inspired dry adhesives (*27-32*), frog- (*33*) and insect-inspired (*34*, *35*) adhesives, and octopus-inspired suction adhesives (*36*, *37*). However, there still remain scientific and technological challenges for successfully mimicking robustness, versatility and strength of biological adhesion systems on real-world surfaces; for example, it is challenging to mimic complicated hierarchical nano-hairs of gecko foot-hairs with conventional microfabrication techniques to achieve high interfacial bonding strength on rough surfaces. In case of the frog- or insect-inspired capillary adhesion, residues from the adhesives as well as the need of continuous supply of liquids on the contact interface significantly limit its real-world applications. On the other hand, suction mechanisms have been widely used for centuries and represent one of the most successful adhesion systems in our daily lives. Yet, its practical applications are mainly focused on flat and smooth surfaces unlike octopi, due to difficulties in achieving secure air-tight sealing on rough and irregular surfaces.

Various attempts have been made to achieve robust suction on rough and irregular surfaces. For example, adhesive surfaces with distributed micro-suction cups have been of interest for conformal contact to a wide range of rough surfaces and human skin in dry or wet conditions (*38*, *39*). The use of micro-suction cups, dividing a single suction area into numerous small gripping areas, lowers the probability of sealing failure propagating from one place to entire contact interface, which enhances the robustness of attachment. Takahashi *et al*. fabricated a vacuum gripper covered with multiple membrane-based suction cups, demonstrating gripping of irregular objects with a step or a groove in dry air, oil, and water (*40*). Baik *et al*. micro-machined an elastomer surface with suction cups inspired by a structure of dome-shaped protuberance inside the suckers of *octopus vulgaris*. The protuberance inside the sucker absorbs excessive liquid on the contact interface, achieving a high pressure differential when engaged to various surfaces in both dry and wet conditions, even on rough and soft skin (*41*). Another approach is to control the interfacial contact area using active components. Zhakypov *et al*. developed an origami-based reconfigurable suction gripper that can actively change shape and size of the suction cup with respect to the object's geometry by actuating artificial muscle wires made of shape memory alloys (*42*). Although these previous approaches proposed various solutions for the grand challenge of developing fail-safe suction mechanisms on rough and irregular real-world surfaces, many limitations still remain on their proposed solution. For example, the use of active components in reconfiguring the suction cups complicates the overall gripping system. Also, such active systems (*13*, *42*) often involve heating to change the gripping status, requiring a relatively long time delay between picking and releasing of the objects, compared to passive systems. Furthermore, bioinspired adhesive surfaces with micro-suction cups (*39*, *41*) require a high preload stress when engaging with an object, posing potential damage issues in grasping fragile objects or living tissues.

In contrast to the previous studies relying on active components or micromachining of the gripper surface, here we report a fully soft suction gripper with no micropatterning, capable of achieving robust attachment on a broad range of three-dimensional (3D) rough object surfaces (**Fig. 1**). The proposed soft suction gripper (SSG) applies suction on the contact interface via a flat non-patterned elastic membrane attached to the soft gripper body. We discovered that the flat membrane improves suction on both smooth and rough surfaces, compared to conventional suction cups without interfacing a membrane. In particular, we found that the SSG passively deforms during the gripping process in a way that the interfacial suction cavity expands proportional to the applied pulling load, resulting in an increased suction force. While capable of conforming to a wide range of real-world 3D surfaces, as well as fragile and deformable objects (**Fig. 1B, Movie S1**), the passive deformation of the SSG depending on the applied load (the object weight reported in **Fig. 1**) can significantly reduce the use of active sensing and sophisticated closed-loop control in robotic grasping (**Fig. 1A**). Furthermore, the structural and material simplicity of the gripper architecture has a potential for a further system-level integration with other soft robotic peripherals, such as robotic fingers and arms, using conventional additive manufacturing techniques (*43-46*). This will provide additional functionalities to soft robotic systems without relying on a sophisticated manufacturing process or complex control systems.

In the following sections, first we discuss the design concept of the SSG, followed by the adhesion mechanism of the entirely soft gripper design, enabling adhesion to a broad range of real-world rough surfaces. Microscopic observations of the adhesion experiments coupled with pressure measurements reveal that the SSG could generate a variable negative pressure differential at the contact interface depending on the applied load, allowing its membrane to conform to a surface morphology with a high level of roughness. Experimental observations on pull-off force measurements between the SSG and a conventional suction cup show that the adhering membrane does not only seal the gap at a contact interface to achieve a conformal contact to a rough surface, but also achieves a high interfacial fracture strength by suppressing unfavorable deformation of the gripper body. Adhesion tests of different membrane materials reveal a trade-off between adaptability and adhesion performance depending on the membrane's elastic modulus, implying an optimal design for various rough surfaces in achieving enhanced suction. Using numerical simulations of the SSG's membrane, we discuss design strategies for improved adaptability to a higher level of roughness to solve diverse challenges in soft robotic grasping.

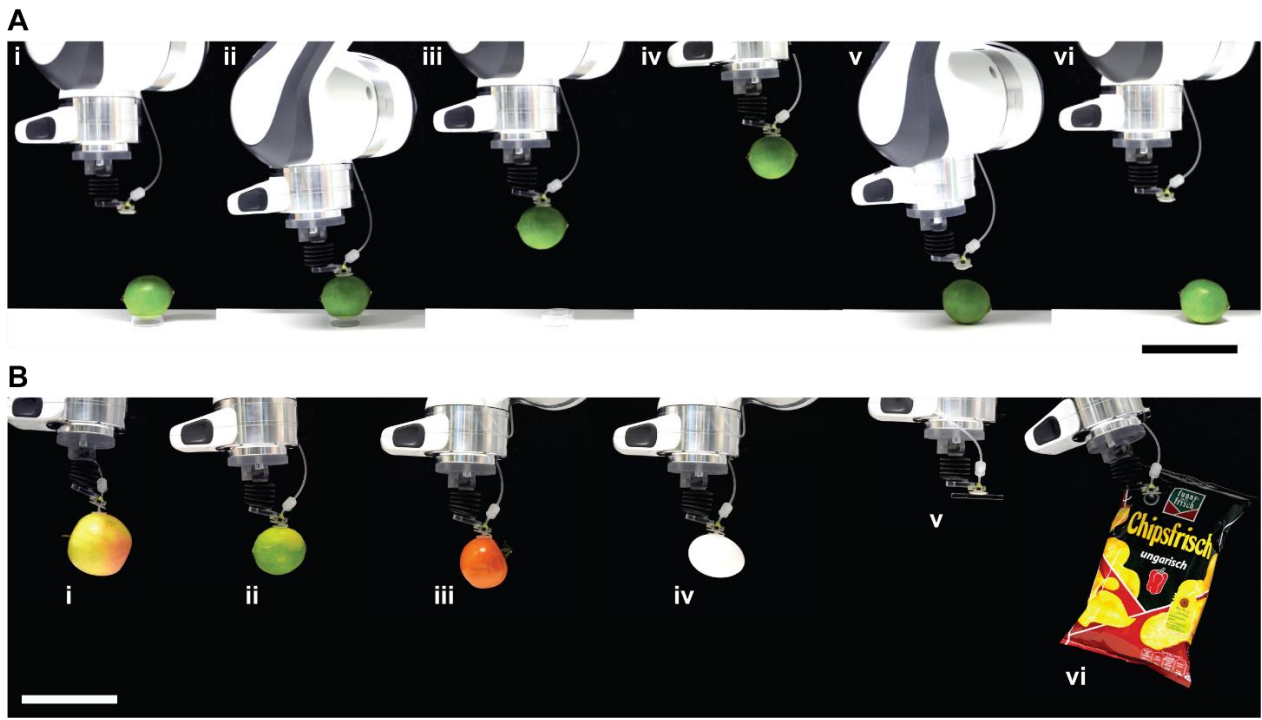

**Fig. 1 | Pick-and-place manipulation demonstration of example rough and irregular real-world objects using the proposed fully soft robotic gripper.** (**A**) Video snapshots of the proposed soft suction gripper attached onto a 6 degrees-of-freedom (DOF) robotic arm (Franka Emika, Panda Research, Germany) manipulating a lime with open-loop control of the gripper. **i**: approaching, **ii**: engaging, **iii**: picking, **iv**: transporting, **v**: releasing, and **vi**: retracting. (**B**) The soft suction gripper holding various objects with a broad range of surface roughness, fragility, shape and weight. **i**: apple (107 grams), **ii**: lime (53 grams), **iii**: tomato (77 grams), **iv**: egg (63 grams), **v**: rectangular bar (2 grams, 5 mm in width, PMMA), and **vi**: plastic bag (53 grams). Scale bars indicate 100 mm.

## Results

### Design of the soft suction gripper

**Fig. 2A** shows a cross-section computer-aided design (CAD) image of the proposed soft suction gripper (SSG). As shown in **Fig. 2B**, the SSG develops a suction force at the contact interface via the flat membrane (FM, 1 in **Fig. 2A**), with respect to a passive deformation of the gripper body (GB, 2 in **Fig. 2A**), when pulled from the contact surface. Here, we define a pressure differential ($\Delta P$) to be a subtraction of the atmospheric pressure ($P_{atm}$) from an internal pressure ($P_i$). The SSG has an overall size of 18 mm in diameter, and a 200-µm-thick flat membrane made out of various off-the-shelf soft elastomers, such as Ecoflex$^{TM}$ (Smooth-on Inc.) and polydimethylsiloxane (PDMS, Sylgard$^{TM}$ 184, Dow Chemical, Co., Ltd.). In the following, we name an SSG with a flat membrane made out of Ecoflex$^{TM}$ 00-50 as 'SSG-Eco50'. During the gripping process, a cotton filter (component 3 in **Fig. 2A**) prevents the membrane from being sucked into the tubing (component 5 in **Fig. 2A**), which can cause a plastic deformation. Under a negative pressure differential ($\Delta P < 0$), the adhering membrane creates a volume called 'suction cavity' at the center of contact interface (6 in **Fig. 2B**), while the rest of membrane seals the contact interface by the collapsed gripper body. Note that the size of the suction cavity varies depending on the applied pulling load, as well as a presence of an initial air pocket.

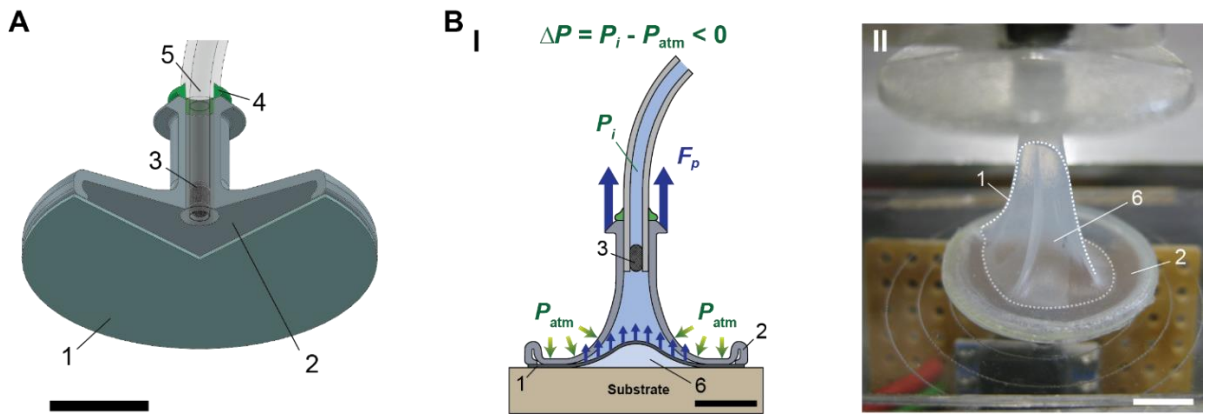

**Fig. 2 | Architecture of the soft suction gripper (SSG). (A)** A cross-section of 3D assembly of the SSG. 1: flat membrane (FM), 2: gripper body (GB), 3: filter, 4: vinylsiloxane, 5: tubing. **(B)** A schematic image of the SSG, being pulled from a contact surface (**I**), and a corresponding photographic image when engaging to a glass surface with an initial air pocket in a hole for measuring the interfacial pressure (**II**). 6: a suction cavity between the adhering membrane and the substrate. A dashed line in **II** shows the outer boundary of the suction cavity. Scale bars indicate 5 mm.

A pull-off experiment on a smooth flat glass surface with a scratch in **Fig. 3** enables visualization of how the SSG adapts to high surface roughness with a soft, deformable membrane. As shown in **Fig. 3A, B**, and **Movie S2**, the SSG is first brought down in contact to the scratched substrate with a preload ($F_l$) (**I**). When removing the air inside the gripper, a negative pressure differential ($\Delta P$) inside the gripper body pulls the flat membrane towards the gripper body, creating a suction cavity (6 in **Fig. 3A-III**, and **3B-III**) at the contact interface (**II**). In the formation of the suction cavity, the ambient air is sucked into the contact interface through the scratch (8 in **Fig. 3A-III**, **3B-II**, and **3D**). During this process, static pressure inside the air channel created by the scratch decreases due to the air flow, having the membrane sucked in and seal the gap (**Fig. 3D**). Therefore, while the center of the membrane is pulled inwardly towards the gripper body to create the suction cavity, the outer part of the membrane on the scratch is pulled in the opposite direction towards the air channel to seal the scratch. During retraction of the SSG, the soft gripper body deforms with different intrinsic curvatures under a uniaxial loading, evolving into a tripod-shaped structure (*47*), as observed in **Fig. 2B-II**. The suction cavity expands from the center towards the edge of the gripper with respect to the increase in body deformation, as seen in **Fig. 3A-III**, **IV**. When reaching to the maximum pull-off load ($F_p$), those cone-shaped deformation (9 in **Fig. 3C-I**) reaches to the edge of the gripper, peeling the adhering membrane off from the contact interface, and breaking the sealing on the contact interface (**V**), as shown in **Fig. 3C** and **Movie S3**. The suction cavity will disappear, as the pressure differential at the contact interface can no longer be maintained due to leakage. **Fig. 3E** shows a representative reaction force ($F_r$) measurement of the SSG with its corresponding $\Delta P$. When approaching to a surface, the gripping system moves down until the contacting force (positive $F_r$) reaches 0.5 N as a preload ($F_l$). After a predetermined contact time of 40 s, the gripper is slowly unloaded at a speed of 100 μm·s$^{-1}$ to minimize viscoelastic effects on the adhesion. The $\Delta P$ at the moment of retraction is defined as an applied pressure ($P_a$). During the retraction, $F_r$ reduces due to the suction at the contact interface, and the maximum value is defined as a pull-off force ($F_p$). **Fig. 3F** shows $F_p$ on glass surfaces with different radii, depending on $P_a$. The results show a monolithic increase and convergence in $F_p$ with respect to a decrease in $P_a$ for all tested radii of curvature, demonstrating that our fully soft gripper can conform to a wide range of 3D geometries to exert a high suction force.

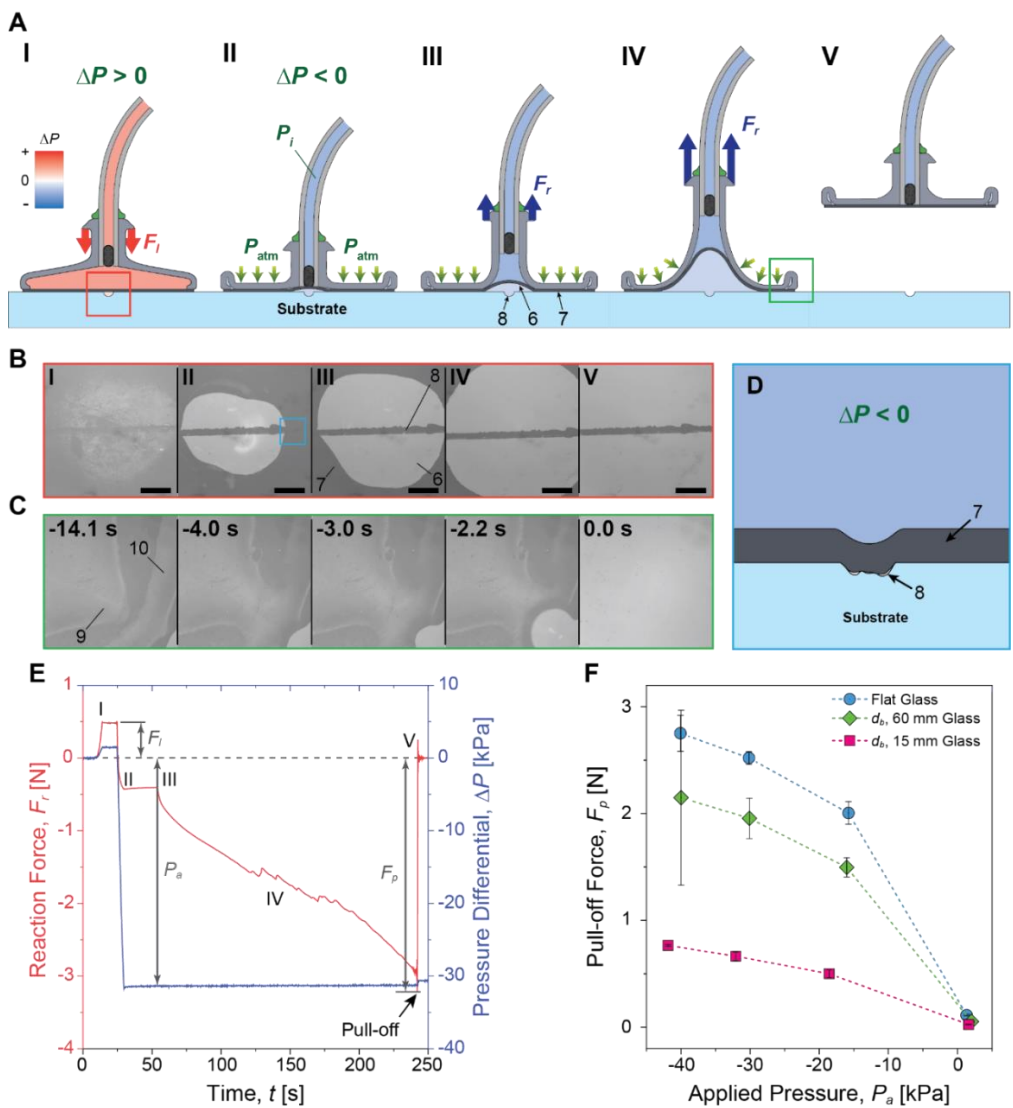

**Fig. 3 | Contact interface visualization of the soft suction gripper membrane on various glass surfaces.** (**A**) A schematic procedure of a pull-off experiment of the SSG on a flat glass surface with a scratch. Preloading (**I**), engaging the substrate with a negative pressure differential (Δ$P$) (**II**), initiating the retraction (**III**), pulling-off the gripper (**IV**), and detachment (**V**). 6: a suction cavity at the contact interface, 7: the area of adhering membrane in contact, and 8: scratch. (**B**) Microscopic images of the scratched flat glass at the center of the membrane (the red inset in **A-I**), corresponding to the experimental procedure from **I** to **V** in **A**. (**C**) Microscopic images of the edge of the gripper at the contact interface (the green inset in **A-IV**), when pulled off. 9: tripod-shaped deformation, 10: edge of the gripper. (**D**) A schematic cross-section image of the contact interface nearby the scratch (the blue inset in **B-II**). (**E**) Profiles of reaction force ($F_r$) and pressure differential (Δ$P$) inside the suction gripper on a flat glass surface with respect to elapsed time ($t$). **I** ~ **V** correspond to the experimental procedure in **A**. (**F**) Pull-off forces ($F_p$) on different radii of glass sphere surfaces depending on the applied pressures ($P_a$). Each point indicates an average of five measurements with corresponding error bars of ±1 standard deviation (SD). Scale bars in **B** and **C** indicate 1 mm.

## Interfacial suction for robust attachment on rough surfaces

**Fig. 4** shows advantages of the proposed SSG over conventional suction cup designs in achieving improved adhesion on rough surfaces. We first replicated various real-world rough surfaces using a double molding process with a clear epoxy polymer (EpoxAcast™ 690, Smooth-on Inc.) as shown in **Fig. S1** (see Materials and Methods for details). The polymeric replicas allow controlled adhesion testing for a wide variety of distinctive roughness features on different surfaces, while keeping the surface chemistry constant. **Fig. 4A** shows the 3D topology of the samples from rough surface (RS) 1 to RS 4, corresponding to **Fig. 4A-I** to **4A-IV**, respectively. As shown in **Fig. 4A-I** and **4A-III**, RS 1 and RS 3 are relatively smooth in the micro scale, while showing wavy surface features up to 46 μm in the maximum height difference (**Fig. 4A-III**). Meanwhile, RS 2 shows relatively flat and microscopic roughness with only 14 μm in the maximum height difference (**Fig. 4A-II**). **Fig. 4A-IV** shows RS 4 has both micro and macroscopic roughness up to 51 μm of the maximum height difference. **Fig. 4B** shows root-mean-square (RMS) roughness ($R_q$) of all surface samples, depending on cut-off wave length ($\lambda_c$). While $R_q$ with a short $\lambda_c$ (i.e., $\lambda_c = 3$ μm) represents the surface roughness in micro scale, $R_q$ with a long $\lambda_c$ (i.e., $\lambda_c = 100$ μm) shows more macroscopic roughness. In **Fig. 4B**, SG stands for Smooth Glass, while RS 5 is a replica of a 2000 grit sand paper. Similar to our observation in **Fig. 4A**, **Fig. 4B** shows that RS 2 has a high $R_q$ in micro scale, compared to RS 1 and RS 3. In case of RS 4, although having a similar $R_q$ to RS 3 in macro scale, the surface shows 7.4 times higher microscopic $R_q$. Among all tested surfaces, RS 5 shows the highest $R_q$ in both micro and macro scale, 3 times higher $R_q$ in micro scale and 1.8 times higher $R_q$ in macro scale, respectively, compared to those of RS 4.

**Fig. 4D** shows a series of microscopic images of the SSG with a flat membrane made out of Ecoflex™ 00-50 (SSG-Eco50), adhering to RS 1. **Fig. 4D-I** to **Fig. 4D-IV** match with experimental procedures in **Fig. 3B-I** to **Fig. 3B-IV**, respectively. When first bringing down in contact to the RS 1 in **Fig. 4D-I**, a significant portion of the flat membrane shows no conformation to the surface topology, indicated as a bright area. As a negative pressure differential (Δ$P$) is applied to the inside of the gripper body in **Fig. 4D-II**, in contrary, almost the entire surface area of the membrane conforms to the rough surface, shown as a dark area (7 in **Fig. 4D-III**). As explained in **Fig. 3D**, the creation of the suction cavity at the center of the SSG reduces static pressure at the contact interface. The deformable, soft membrane adapts to the majority of roughness by the negative Δ$P$ at the interface, while a small amount of trapped air is coalesced in the suction cavity at the center of the SSG (6 in **Fig. 4D-II**). As the SSG is retracted from the contact surface, **Fig. 4D-III**, **4D-IV**, and **Movie S4** show that small microcavities trapped at the contact interface grow with respect to the increase in the amount of negative $\Delta P$ (13 in **Fig. 4D-IV**).

**Fig. 4E** shows that pull-off forces ($F_p$) of the SSG on SG and RS 1 ~ RS 4 have 3.8 times higher suction in average, compared to the gripper without a membrane. Note that both the SSG-Eco50 and the gripper body without a membrane readily fail on RS 5 due to the high roughness. Under a high pulling load as shown in **Fig. 4C**, the gripper body experiences a high tensile force at the gripper edge acting toward the center of gripper (blue arrow). In case of the gripper body without a membrane, only a small portion of the gripper body remains in contact to the rough surface under the high pulling load (**Fig. 4C-I**). Although a small amount of interfacial friction between the gripper body and the rough surface exists (pink arrow), the gripper body instantaneously fails upon slipping of the gripper edge caused by the high tensile force. In case of the SSG, on the other hand, the area of contact interface between the adhering membrane and the rough surface remains constant, no matter how much the gripper body deforms by the high pulling load (**Fig. 4C-II**). This high interfacial friction by the membrane suppresses lateral deformation of the SSG at the contact edge, resulting in high $F_p$. Considering the maximum theoretical pull-off force ($F_p|_{max}$) is estimated by $P_a \pi d^2/4$ ($P_a$: applied pressure, $d$: gripper diameter), SSG-Eco50 could reach up to 36% of $F_p|_{max} = 7.7$ N, whereas the gripper body without a membrane ended up achieving only 10% of $F_p|_{max}$, showing the SSG's enhanced gripping robustness on a broad range of rough surfaces, compared to the conventional suction cup designs.

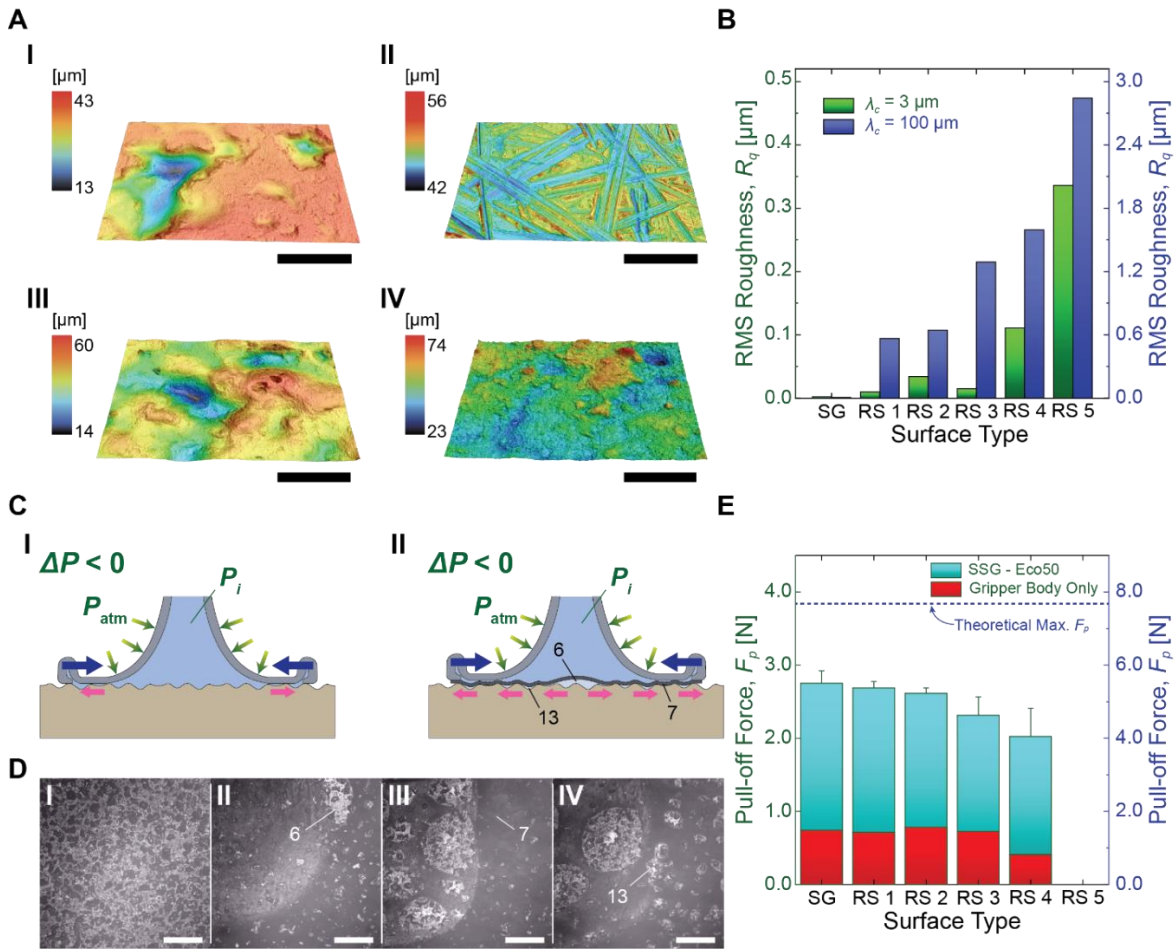

**Fig.4 | Adaptability of the soft gripper on rough surfaces.** (**A**) Topographical 3D-scanned images of rough surfaces (RS) measured by a laser confocal microscope. (**I**): RS 1, (**II**): RS 2, (**III**): RS 3, and (**IV**): RS 4. Scale bars indicate 200 µm. (**B**) Root-mean-square (RMS) roughness ($R_q$) of the tested surfaces, when cut-off wave length ($\lambda_c$) is 3 µm (green) and 100 µm (blue). (**C**) Schematic

illustrations of attachment of (**I**) the soft gripper body, and (**II**) the gripper on a rough surface. Blue arrows indicate tensile forces acting on the gripper edge, while pink arrows show friction forces between the gripper and the contact interface. 6: a cavity at the contact interface, 7: the membrane area in contact, and 13: microcavities at the contact interface due to roughness. (**D**) Microscopic images of the contact interface on the RS 1 at the center of the membrane, corresponding to the experimental procedure from **I** to **IV** in Fig. **3A**. Note that the microcavities (13 in **A-II**) expand as the gripper is retracted (compare **III** and **IV**). Scales indicate 1 mm. (**E**) Pull-off force ($F_p$) comparison between the SSG-Eco50 (**A-II**) and the gripper body (**C-I**) on various rough surfaces (RS 1 ~ RS 5), and on a smooth glass (SG). The theoretical maximum of pull-off force ($F_p$) follows the scale of a y-axis on the right.

**Fig. 4D** shows the growth of microcavities, during the retraction of the SSG, implying that interfacial pressure differential changes with respect to motion of the gripper. In **Fig. 5**, we measured the suction at the contact interface across the adhering membrane, depending on retraction distance ($z_r$) of the SSG. Here, we call interfacial pressure differential simply interfacial pressure ($\Delta P_i$) in the further discussion. As shown in **Fig. 5A-D**, a pressure sensor (HSCSANN600MDAA5, Honeywell International Inc., 12 in **Fig. 5A-B**) is attached on a smooth glass slide through a 500-µm-diameter connecting hole (11 in **Fig. 5A-D**). **Fig. 5E** shows profiles of $\Delta P_i$ with respect to $z_r$, depending on different sensor offsets ($d_o$) from the center of the SSG. $\Delta P_i|_{\text{III}}$ refers to an interfacial pressure at $z_r = 0$ mm, while $\Delta P_i|_{\text{pull}}$ is an interfacial pressure when the SSG reaches the pull-off force ($F_p$).

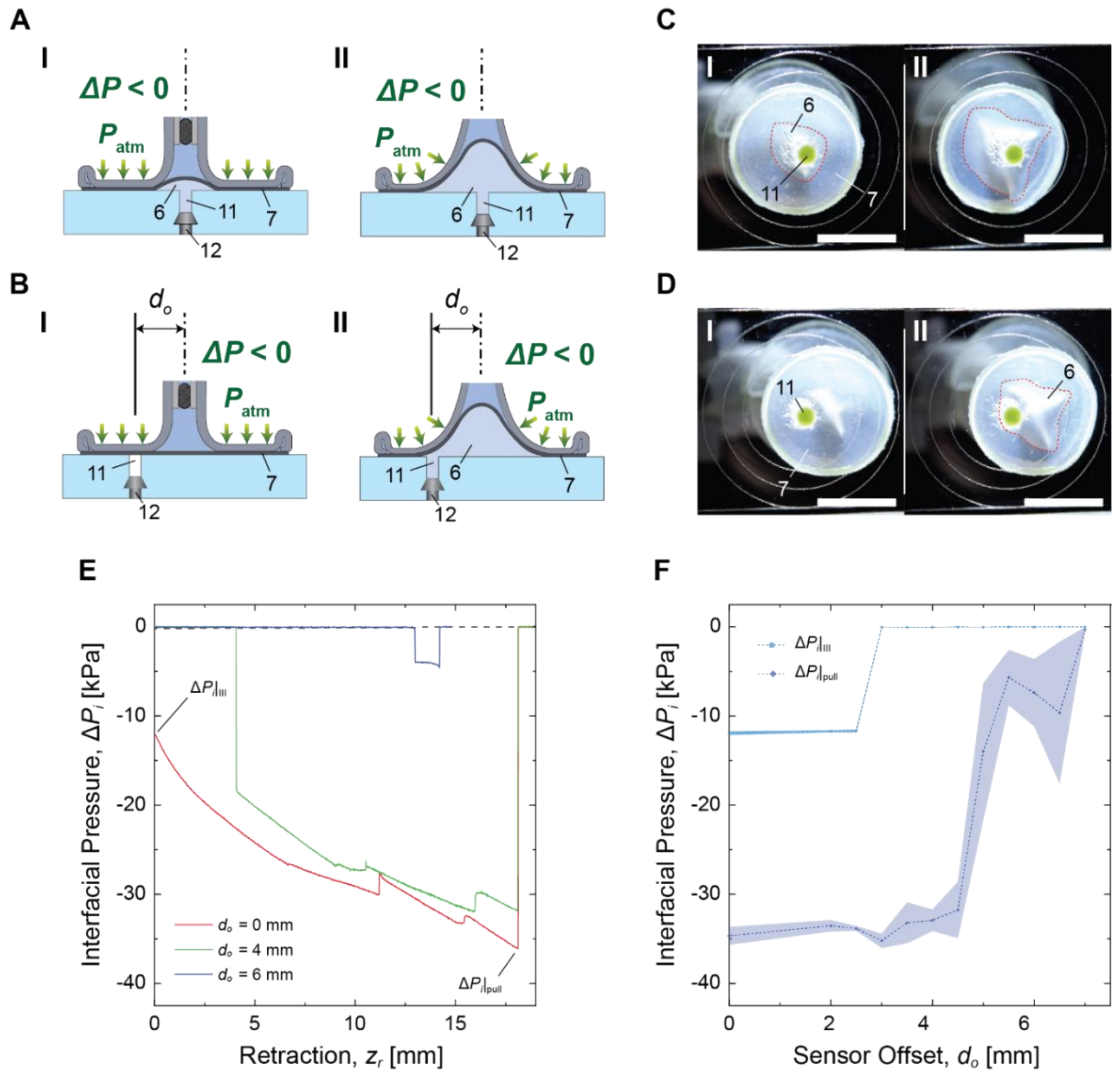

**Fig.5 | Measurements of the interfacial pressure differential of the soft gripper.** (**A**) Schematic illustrations of the gripper whose center is aligned with the pressure sensor, when initiating the retraction (**I**), and pulling-off the gripper (**II**). 6: a cavity at the contact interface, 7: the area of membrane in contact, and 11: connecting hole, 12: pressure sensor (**B**) Schematic illustrations of the gripper when the interfacial pressure is measured off-center with an offset $d_o$ of the pressure sensor, when initiating the retraction (**I**), and at the pull-off force right before detachment (**II**). (**C**) Photographic images of contact interface of the SSG-Eco50 when the interfacial pressure is measured at the center, when initiating the retraction (**I**), and at the pull-off force right before detachment (**II**). The red dashed line indicates the suction cavity (6 in **A** and **B**). (**D**) Photographic images of contact interface of the SSG-Eco50 when the interfacial pressure is measured off-center by $d_o = 4$ mm, when initiating the retraction (**I**), and at the pull-off force before the detachment (**II**). (**E**) Representative profiles of interfacial pressure ($\Delta P_i$) of the SSG-Eco50 at the center, depending on the gripper offsets ($d_o$) with respect to retraction ($z_r$). $\Delta P_i|_{\text{III}}$ indicates the interfacial pressure when initiating the retraction, corresponding to the procedure in **Fig. 3A-III**. $\Delta P_i|_{\text{pull}}$ indicates the interfacial pressure when the gripper reaches at the pull-off force ($F_p$). (**F**) Profiles of interfacial pressure ($\Delta P_i$) of the SSG-Eco50 with respect to the gripper offsets ($d_o$), when retraction ($z_r$) is at $\Delta P_i = \Delta P_i|_{\text{III}}$ and $\Delta P_i = \Delta P_i|_{\text{pull}}$. The shaded areas indicate standard deviation, while the dashed lines are the arithmetic average of five measurements. Scales in **C** and **D** indicate 10 mm.

In case of $d_o = 0$ mm (**Fig. 5A, C**), applied pressure ($P_a$) inside the SSG creates a small suction cavity (**Fig. 5C-I**) with an interfacial pressure $\Delta P_i|_{\text{III}} = -11.9$ kPa, when $z_r = 0$ mm. The size of suction cavity passively grows with respect to $z_r$, while $\Delta P_i$ decreases until it reaches $\Delta P_i|_{\text{pull}} = -36.1$ kPa, when the SSG is pulled off. On the profile curve, small upsurges on $\Delta P_i$ occur when deformation of the gripper body introduces trapped air bubbles to the suction cavity, which causes following reposition of the SSG (**Movie S5** and **Movie S6**). In case of $d_o = 4$ mm, and $d_o = 6$ mm (**Fig. 5B and 5D**), on the other hand, $\Delta P_i$ initially begins at 0 kPa, as the suction cavity has not reached out to the pressure sensor that is covered by the adhering membrane. (**Fig. 5B-I**). Note that the suction cavity is not visible at the smooth glass interface (**Fig. 5B-I** and **Fig. 5D-I**), until the expanding suction cavity reaches the pressure sensor and obtains the trapped air from the connecting hole (**Fig. 5B-II** and **Fig. 5D-II**). There is a sudden pressure drop in the profile of $\Delta P_i$, as soon as the suction cavity reaches to the pressure sensor (**Fig. 5E**). In case of $d_o = 6$ mm, $\Delta P_i|_{\text{pull}}$ could not reach up to a negative $\Delta P_i$ as high as that of $d_o = 0$ mm and $d_o = 4$ mm, since the connecting hole is reclosed rapidly by the gripper body after the air inside the hole is transferred to the suction cavity. According to **Fig. 5F**, the smallest size of suction cavity is approximately 5 mm in diameter, while the SSG can achieve its maximum payload for objects larger than 10 mm in diameter.

**Fig. 6** shows an important trade-off between pull-off force ($F_p$) and surface adaptability, depending on the softness of its flat membrane. Three SSGs are fabricated with membranes made out of Ecoflex™ 00-10, Ecoflex™ 00-50, and Sylgard™ 184 (PDMS). Young's modulus ($Y$) and elongation at break ($\varepsilon_{\text{max}}$) of these elastomers are reported as follows; Ecoflex™ 00-10: $Y = 50$ kPa, $\varepsilon_{\text{max}} = 573\%$, Ecoflex™ 00-50: $Y = 100$ kPa, $\varepsilon_{\text{max}} = 860\%$, and PDMS: $Y = 2.4$ MPa, $\varepsilon_{\text{max}} = 135\%$, respectively (*48*). **Fig. 6A** shows that the SSG with a stiffer, and less stretchable membrane generally provides a higher $F_p$ on a smooth glass surface at a lower $P_a$. For example, the SSG with a PDMS membrane (SSG-PDMS) could exert up to $F_p = 5.4$ N with $P_a = -30.2$ kPa, while the SSG with an Ecoflex™ 00-10 membrane (SSG-Eco10) could reach up to $F_p$ of 3.4 N that is less than the SSG-PDMS, but with a higher $P_a$ of -15.7 kPa. Although the SSG with a softer membrane could not achieve the $F_p$ as high as that of the SSG with a stiffer membrane, **Fig. 6B**, **C** show a clear benefit of using the softer membrane to conform to rough surfaces. While the SSG-Eco10 could only provide 63 % of the maximum $F_p$ of the SSG-PDMS on the smooth glass, **Fig. 6B** shows the SSG-Eco10 could adhere to all of the tested rough surfaces from RS 1 to RS 4 with an average $F_p = 2.5$ N, while the SSG-PDMS failed to cling onto surfaces with high roughness (RS 3 and RS 4), as shown in **Fig. 6C**. The above results remind a similar trend in adhesion of gecko-inspired micro-fiber adhesives reported in our previous work (*49*), that micro-fibers made out of a stiff and non-deformable polymer could achieve a significantly high interfacial fracture strength for a specific, optimal tip geometry, while micro-fibers with a soft and deformable polymer could exert a less fracture strength at the contact interface, but for a more broad range of tip geometries. Although there does not exist homogeneous elastomer for the membrane that shows both softness and non-stretchability at the same time, we expect a heterogenous composite membrane that is made out of a non-stretchable material (i.e., fabrics) covered with a soft elastomer may achieve the above favorable properties to exert improved suction forces on a wide range of real-world surfaces with high roughness.

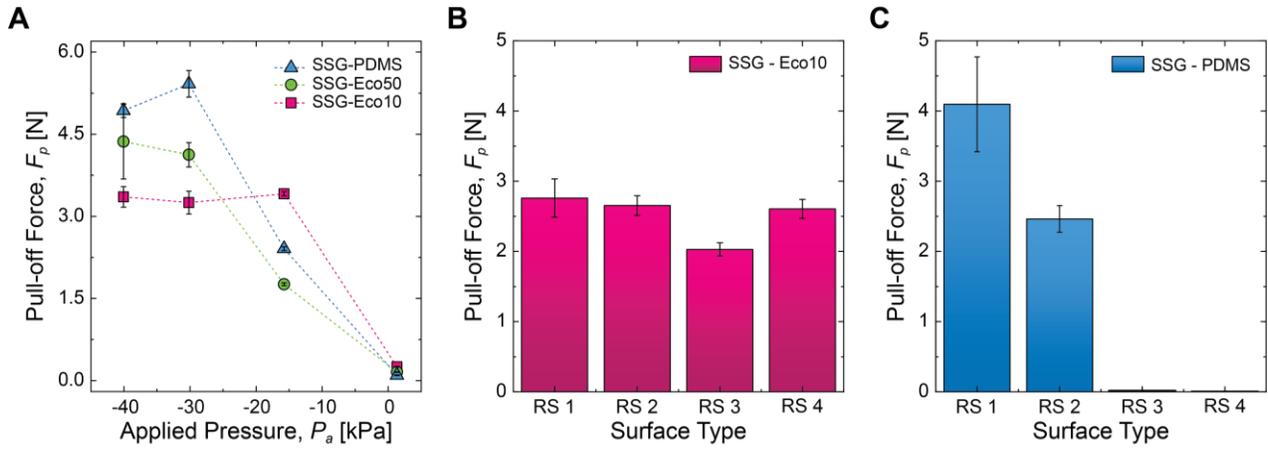

**Fig. 6 | Trade-off between maximal suction force and surface conformation/adaptation capability.** (A) Pull-off force ($F_p$) of the soft gripper with respect to applied pressure ($P_a$), depending on elastomeric materials of the flat membrane. PDMS: polydimethylsiloxane, Eco50: Ecoflex™ 00-50, and Eco10: Ecoflex™ 00-10. (B) Pull-off force ($F_p$) of the SSG-Eco10 on different rough surfaces from RS 1 to RS 4. $P_a$ is -30.2 kPa for all measurements. (C) Pull-off force ($F_p$) of the SSG-PDMS on different rough surfaces from RS 1 to RS 4. $P_a$ is -30.2 kPa for all measurements. Error bars indicate standard deviation of five measurements.

## Numerical simulation for guiding the design of the soft suction gripper

**Fig. 7** shows effects of various design parameters of the SSG, depending on surface roughness and pressure differential ($\Delta P$). **Fig. 7A-I** depicts the geometrical parameters including the membrane thickness ($t$), roughness parameters (cut-off wavelength ($\lambda_c$) and amplitude of the wave ($R_z$)), and applied $\Delta P$. The rough surface is represented by sinewaves in two dimension (2D), assuming that the wavelength (here coincide with $\lambda_c$) of the sinewave determines the dominant feature size of the rough surface. Note that $R_z$ can be estimated as a function of RMS roughness ($R_q$) and $\lambda_c$. Given those parameters, numerical simulation calculates the minimum $\Delta P$ required to seal the gap between a flat membrane and the sinusoidal-shaped substrate as shown in **Fig. 7A-II**.

According to experimental results in **Fig. 6B**, **C**, the critical $R_q$ on which the SSG-Eco10 fails to adhere is estimated as $R_q = 0.34$ μm at $\lambda_c = 3$ μm, or $R_q = 2.8$ μm at $\lambda_c = 100$ μm, while the SSG-PDMS will readily fail on a rough surface with $R_q = 0.03$ μm at $\lambda_c = 3$ μm, or $R_q = 1.3$ μm at $\lambda_c = 100$ μm, respectively. As a prerequisite for an intimate contact, the flat membrane needs to seal both micro- and macro-sized roughness with the applied $\Delta P$, which can be described as $R_q$ at $\lambda_c = 3$ μm, and $R_q$ at $\lambda_c = 100$ μm, respectively.

**Fig. 7B**, **7C** show the calculated $\Delta P$ of soft and rigid membranes respectively, for the membrane to fully conform to the sinusoidal-shaped roughness, depending on $\lambda_c$ and $R_q$. The results in both graphs show that $R_q$ at a smaller $\lambda_c$ requires a higher $\Delta P$ than $R_q$ at a greater $\lambda_c$, which indicate that micro-sized roughness is more difficult to be adapted than macro-sized roughness. Furthermore, curves in **Fig. 7B** have smaller slopes than those of curves in **Fig. 7C**, as the soft membrane possesses better conformability than the rigid membrane. At $\Delta P = 30$ kPa, which is a pressure differential applied in the experiments (**Fig. 6**), **Fig. 7B** shows that the soft membrane made out of Ecoflex™ elastomer can seal the all macro-sized roughness calculated up to $R_q = 3.0$ μm with the given $\Delta P$, while still experiencing difficulties in conforming to micro-sized roughness greater than $R_q = 0.32$ μm (i.e., RS 5), which well matches with our experimental observation (**Fig. 6B**). In case of the membrane made out of PDMS (**Fig. 7C**), on the other hand, the numerical results predict the rigid membrane will fail on a surface with $R_q$ greater than $R_q = 0.04$ μm at $\lambda_c = 3$ μm, and $R_q = 0.38$ μm at $\lambda_c = 100$ μm, implying that the membrane fails on RS 3, RS 4, and RS 5 in **Fig. 6C** due to its high roughness in macro-scale up to $R_q = 1.3$ μm at $\lambda_c = 100$ μm (**Fig. 4B**).

Numerical analyses in **Fig. 7D**, **E** suggest how to further improve the adaptability of a flat membrane with various design parameters, such as Young's modulus ($Y$) and membrane thickness ($t$). **Fig. 7D** shows predictions on the minimum $R_q$ of the Ecoflex$^{TM}$ membrane to seal the roughness in micro-scale ranging from $R_q = 0.0$ µm to $R_q = 0.4$ µm at $\lambda_c = 3$ µm, depending on various $t$. The results suggest that thickness of the soft membrane needs to be significantly lowered down to $t = 1.0$ µm, in order to achieve adaptability high enough to adhere on to the roughness of RS 5 with $\Delta P = 30$ kPa. Furthermore, the effect of lowering thickness on improving adaptability will be saturated at $t = 5.0$ µm, suggesting thickness higher above $t = 5.0$ µm will have the similar low adaptability to $t = 200$ µm, as shown in **Fig. 6B**. Such an extremely thin, flat membrane over a large surface area is not realistic, as its structural integrity can easily be compromised under a high loading condition, failing to withstand repetitive gripping tasks. Therefore, a structural reinforcement must be combined with the thin, membranous structure in order to exploit the high surface adaptability with sufficient mechanical stability. More realistic approach can be found in **Fig. 7E**, showing the estimated $\Delta P$ of a 200-µm-thick membrane to seal the same micro-sized roughness, depending on various $Y$. The result shows a clear negative correlation between $Y$ and $R_q$ with a given $\Delta P$. Reducing $Y$ by a half (25 kPa) from the current version of the SSG with Ecoflex$^{TM}$ 00-30, our numerical estimation predicts that the flat membrane can grip all the rough surfaces presented in our experiments, although softness of a membrane might limit the maximum pull-off forces as shown in **Fig. 6A**.

In overall, the numerical simulation studies shown in **Fig. 7** provide a design guideline that surface conformation/adaptability of the SSG can be further improved by reducing thickness or stiffness of the membrane, as long as the membrane can endure the applied loading conditions. Also, the competing relationships among adaptability, mechanical integrity, and pull-off force leave the fact that there would be optimal design parameters for a given target range of surface roughness that the SSG needs to grip. Furthermore, the results in numerical simulation support our discussion in **Fig. 6** that a composite membrane consists of an extremely soft material with a non-stretchable fabric reinforcement can indeed achieve both high pull-off force and better conformability simultaneously, which is a future work.

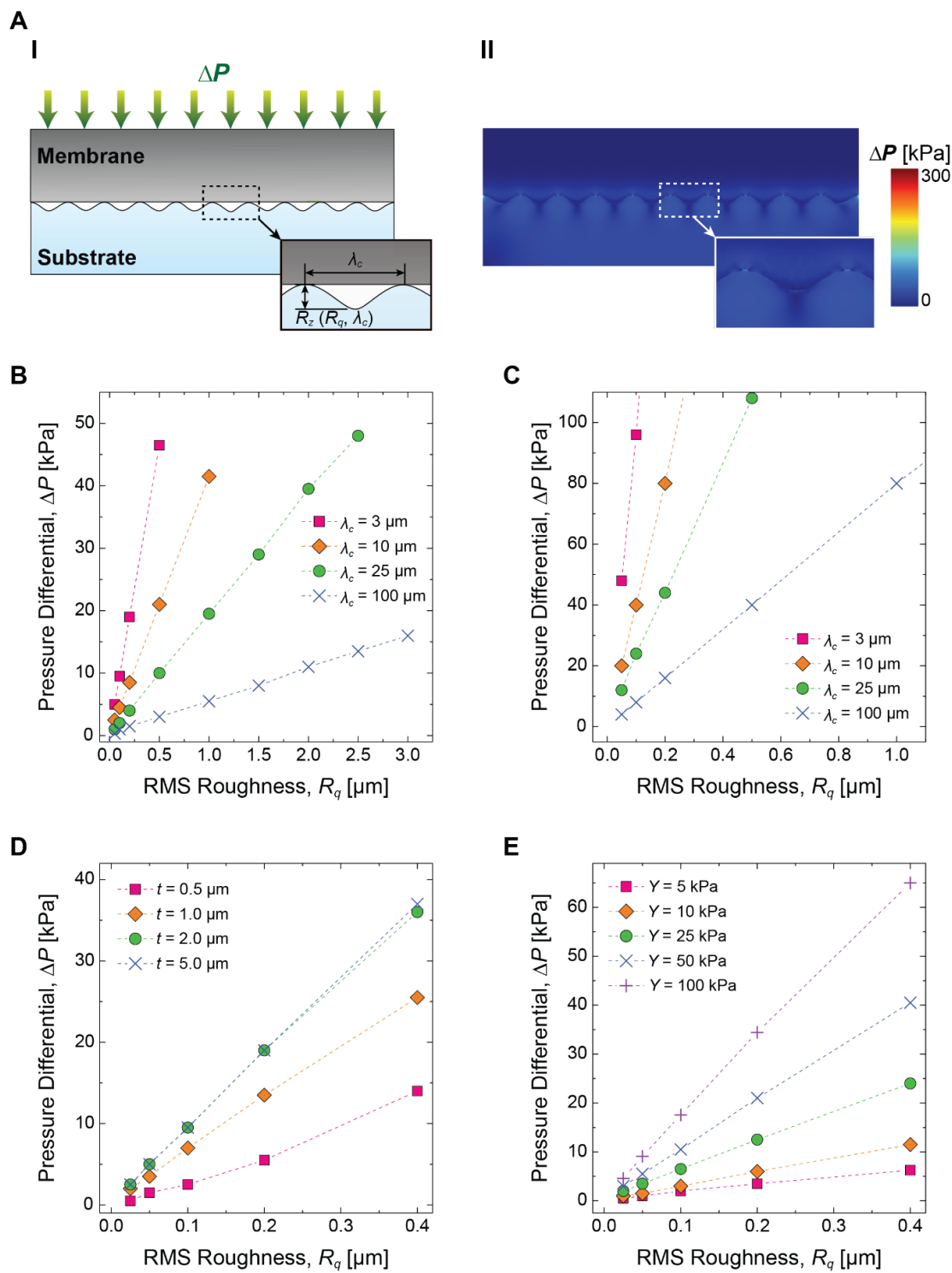

**Fig.7 | Membrane conformability analysis based on design parameters via finite element analysis simulations.** (**A**) Schematic of a simulation sample (I) and the result (II). The membrane is conformed to the rough surface geometry of the substrate by the applied pressure differential, $\Delta P$. (**B, C**) Membrane conformability by the cut-off wavelength, $\lambda_c$, (or the dominant roughness feature size of the surface) with (**B**) Ecoflex$^{TM}$ 00-30 and (**C**) PDMS. (**D**) Membrane conformability by the membrane thickness, $t$, in case of Ecoflex$^{TM}$ 00-30 of 3 μm $\lambda_c$ (the case of PDMS is shown in shown in **Fig. S7**). The result of a membrane thicker than 5 μm is the same as for one of 5 μm, which suggests that the effect of the membrane thickness on the conformability saturates around 5 μm. (**E**) Membrane conformability by Young's modulus, $Y$, of simulated linear materials. A softer material conforms better than a hard material given the same $\Delta P$ and $\lambda_c$.

## Discussion

Experimental and numerical results presented in this work spark a discussion about how to design a soft suction cup as a universal robotic gripper for real-world objects. Suction cups are everywhere; its versatility and usability in our daily lives can easily be found in various locations for different purposes, ranging from holding a car navigation system on a wind shield, attaching a hook on a mirror in the bathroom, to transferring a thin and wide, fragile glass panel in a liquid crystal display (LCD) factory. Those conventional suction cups that consist of rigid plastic parts and hard rubbers are designed to hold a high suction force on smooth surfaces with a small radius of curvature. For robotic grasping, however, we can clearly agree on the need of soft and conformable materials as building blocks of the suction cup, to achieve higher adaptability on irregular geometries and rough surfaces. Unfortunately, the higher adaptability comes with a cost; **Fig. 4E** showed that a suction cup made of a highly deformable silicone without a membrane could not keep an applied negative pressure differential inside the gripper body under a high pulling load, reaching only up to 10% of its theoretical maximum pull-off force. Interestingly, the above result leads us to an encounter with the fundamental trade-off in adhesion between surface adaptability and interfacial fracture strength that we have already experienced, when integrating gecko-inspired micro-fiber adhesives with a soft membrane in our previous work (*50*). Although the gripping mechanisms are different between two cases, the same trade-off mentioned above forces us to sacrifice pull-off force in exchange of surface adaptability by using soft materials. Rather complying with such a trade-off, here we proposed an entirely soft suction gripper that interlaces a soft, flat membrane between the gripper body and an object surface. The proposed soft suction gripper did not only show superior adaptability to 3D geometries with a large radius of curvatures (**Fig. 3F**), but also exhibited high pull-off forces up to 36% of its theoretical maximum (**Fig. 4E**). Our initial concept of having a soft, unstructured membrane at the contact interface can provide both superior adaptability and high fracture strength. Experimental observations in **Fig. 4D** suggest that a suction cavity located at center of the gripper can create an interfacial suction to bring the majority of the membrane into an intimate contact to the object's surface, which contributes to the improved pull-off forces in **Fig. 4E** by stabilizing the entire gripper structure with a friction (**Fig. 4C-II**). In addition to the robust suction and the high adaptability on rough and irregular surfaces, our gripper also shows embedded physical intelligence as a form of passive deformation of the soft gripper body to change the size and strength of the suction cavity (**Fig. 5**), which could simplify manipulation tasks of a wide variety of real-world objects without using sensor input and sophisticated closed-loop control (**Fig 1**).

Obviously, this is not the first work in terms of integrating a membrane structure with a suction cup for the first time. For example, Tsukagoshi *et al*. have developed a ball-shaped casting device covered with a thermoplastic elastomer that can adhere onto a surface of rough wall by suction (*51*). Mazzolai *et al.* have also covered their soft suction cups on an octopus-inspired arm with a membrane for the purpose of isolating the entire suction system from the surrounding environment, and preventing hydraulic leakage for other suction cups, when one of the suckers failed to grip the object surface (*52*). However, to the best of our knowledge, there does not exist a comprehensive research and investigation about advantages of having a soft membrane at the contact interface for improved adhesion on rough surfaces, as well as effects of various design parameters of the soft membrane on suction strength and adaptability. **Fig. 6** reveals that membrane stiffness and stretchability indeed have significant effects on surface conformation/adaptability and pull-off force of the SSG. The results suggest that the SSG with a stiffer membrane can exert an improved suction on a smooth surface, while the SSG with a softer membrane will be suitable for engaging surfaces with high roughness. Using numerical simulations, we have conducted a quantitative investigation on the effects of various design parameters, such as the membrane stiffness, membrane thickness, and surface roughness, on the adaptability of the SSG. **Fig. 7** shows that a thinner and softer membrane can possess enhanced adaptability, as long as the membrane of an SSG can support the applied loading condition. The results shown in Fig. 7 suggest that the trade-off among adaptability, mechanical robustness, and pull-off force must be considered for design of the SSG based on its target applications. There remain several future works to improve the proposed SSG with better performance. For example, as suggested by results in **Fig. 6** and **Fig. 7**, the use of a composite membrane that consists of a non-stretchable backing (i.e., fabric or a thin layer of stiff elastomer) covered with a soft elastomer has a potential to achieve both high suction force and superior adaptability on a broad range of rough surfaces. Furthermore, the soft robotic architecture of the proposed SSG can easily be combined with various functional interfaces developed in previous works, such as electrostatically adhesive surfaces (*53*), crack trapping surfaces (*54*), and micro-suction surfaces (*41*), to produce better performance in achieving robust and strong adhesion. We envision that the simplicity as well as the versatility of the proposed soft suction gripper in this research will serve as a cornerstone in developing universal soft robotic gripping devices that can provide an important functionality for the next generation of interactive robotic systems in the future.

## Materials and Methods

### Fabrication of the soft suction gripper

The soft gripper body was obtained by replicating a negative mold (**Fig. S4**) as previously reported (*50*). In short, a negative mold made out of Ecoflex$^{TM}$ 00-30 (Smooth-On Inc.) was obtained by replicating a 3D-printed composite model of the gripper body (GB). The gripper model was designed with a CAD software and created by a 3D printer (Objet260 Connex, Stratasys Ltd.) using a stiff (VeroClear$^{TM}$) soft (TangoBlack$^{TM}$) material.

The chamber was attached to a small plastic petri dish and a 1:1 ratio of Ecoflex$^{TM}$ 00-30 prepolymer and crosslinker was mixed, degassed, and casted into the petri dish. After curing at room temperature for 6 hours (**Fig. SI 1**) the gripper model was carefully demolded. The negative mold was treated in an oxygen plasma for 2 minutes, fluorosilanized for 1 hour and cured at 90 °C for 30 minutes. The fluorosilanization of the mold allows the subsequent replication process. A thin metal bar was applied to the negative mold (**Fig. SI 1, v**) and a 1:1 ratio of Ecoflex$^{TM}$ 00-50 (Smooth-On Inc.) parts A and B were mixed, degassed and injected into of the negative mold with a syringe. After curing at room temperature for 14 hours, the soft GB was demolded.

The SSG membranes were fabricated of materials with different Young´s moduli (**Fig. S5 A**). A 1:1 ratio of Ecoflex$^{TM}$ 00-50 (Smooth-On Inc.) Parts A and B were mixed, degassed, and casted on perfluorinated silicon wafer. A ca. 200 µm thin layer was created by a bar coater (K-Hand-Coater, Erichsen GmbH & Co. KG) and cured at room temperature for 14 hours and peeled off. The Ecoflex$^{TM}$ 00-10 and Sylgard$^{TM}$ 184 (Dow Chemical Co., Ltd.) membranes were fabricated as described before, but cured at room temperature for 6 hours and cured in a vacuum oven at 90 °C for 1 hour, respectively.

The membrane was bonded to the gripper body with a vinylsiloxane polymer (Flexitime® Medium Flow, Heraeus Kulzer GmbH) (**Fig. S2 B**). The vinylsiloxane precursor was applied on a glass plate and a thin film of ca. 50 µm thickness was created by a film applicator (Multicator 411, Erichsen GmbH & Co. KG). The soft GB was manually inked into the polymer film and placed on a flat membrane. The vinylsiloxane covalently bonds the GB to the membrane after 5 minutes of curing at room temperature. The vinylsiloxane polymer was also used in assembling the connecting tube with the gripper body.

### Fabrication of the rough surface replicas

A vinylsiloxane polymer (Flexitime® medium flow, Heraeus Kulzer GmbH) was applied on a glass slide and a ca. 2 mm film was created (**Fig. S2**). 1 mm thick spacers were attached on the edges of the glass slide. The vinylsiloxane film on top of the glass slide was pressed against a real surface e.g. concrete, steel, ceramic and polymer (**Fig. SI 3**) and cured for 5 min at room temperature (**Fig. S3**). After curing the sample was peeled off the surface. A 10:3 ratio of EpoxAcast$^{TM}$ 690 (Smooth-On Inc.) prepolymer and crosslinker was mixed, degassed, and casted onto the surface replica, followed by another degassing step to ensure optimal replication quality. A glass slide with 1mm thick spacers on the edges was placed on top, serving as a backing layer and cured at room temperature for 48 hours (**Fig. SI 4**). After demolding a positive replica of the rough surfaces was obtained.

## Experimental Setup

The soft gripper was characterized with a customized adhesion setup described before (*50*). In short, the adhesion setup was mounted on an inverted optical microscope (Axio Observer A1, Zeiss) with a video camera (Grasshopper®3, Point Grey Research Inc.). The adhesion was measured by high-resolution load cells (GSO-25, GSO-500, and GSO-1K, Transducer Techniques®), which was attached on a computer-controlled high-precision piezo motion stage (LPS-65 2", Physik Instrumente GmbH & Co. KG) in z-direction. The substrate was placed on a sample holder and moved in *x*-direction by the piezo stage (LPS-65 2", Physik Instrumente GmbH & Co. KG). Fine positions in *x*- and *y*-direction and angular alignment were performed by a manual *xy*-stage (NFP-2462CC, Positionierungstechnik Dr. Meierling) and by two goniometers (M-GON65-U, Newport), respectively. The pressure inside the SSG was controlled by a syringe pump (Legato$^{TM}$ 210P, KDScientific Inc.) and measured with a pressure sensor (HSCSANN600MDAA5, Honeywell International Inc.). The motion of the piezo stages and the data acquisition were performed by a customized code in Linux (Ubuntu$^{TM}$, Canonical Ltd.).

## Finite element modeling (FEM) simulations

The simulation studies were conducted in a commercial FEM software (COMSOL Multiphysics®, COMSOL Inc.). In the simulation, the geometry of the substrate (**Fig. 7A-I**) to yield a targeted RMS roughness value, $R_q$, a sinusoidal surface shape was used. While the wavelength of the sinusoidal wave is given by the specific cut-off wavelength ($\lambda_c$), the amplitude of the wave ($R_z$) is calculated to match the targeted $R_q$ value. Ten waves on the substrate surface were used to test the conformability of the membrane on the substrate. In the simulation, PDMS was used as a rigid membrane, while Ecoflex$^{TM}$ 00-30 was used as a soft membrane, considering similar mechanical properties between Ecoflex$^{TM}$ 00-10 and Ecoflex$^{TM}$ 00-30 elastomers reported in several different previous works (*48*, *55*). To simulate the nonlinear material behaviors of PDMS and Ecoflex$^{TM}$ 00-30, the 5 parameter Mooney-Rivlin hyperelastic model is employed. The pressure applied to the top part of the membrane induces the deformation of the membrane. The pressure is increased iteratively until the gap between the membrane and the substrate closes. The gap is determined to be closed when the maximum distance between the membrane and the substrates goes below 9 nm. During the whole simulation, the boundary condition at both ends were kept frictionless, using roller constraints in COMSOL. The linear material simulation (**Fig. 7E**) is conducted in a same way as described above, except the materials behavior is assumed to be linear. For the numerical stability in the simulation, still the Mooney-Rivlin hyperelastic model is employed while the stress and strain relationship kept linear by fitting Mooney-Rivlin coefficients to the linear line with a slope of a given Young's moduli.

**Funding:** This work is funded by the Max Planck Society.

**Author contributions:** S.S., D.-M.D., and M.S. proposed and designed the research; S.S., D.-M.D., and A.K. performed the experiments; S.S, and D.-M.D. analyzed the experimental data; D.S. performed the numerical analysis; and S.S, D.-M.D., D.S., and M.S. wrote the paper. S.S. and D.-M.D. contributed equally.

**Competing interests:** Max Planck Innovation will file a patent application for the novel gripper device in this study.

Supplementary Materials for

# Soft Robotic Suction Grasping for Rough and Irregular Surfaces

Sukho Song[1,2]*, Dirk-Michael Drotlef[1]*, Donghoon Son[1], Anastasia Koivikko[1,3], and Metin Sitti[1†]

[1]Physical Intelligence Department, Max Planck Institute for Intelligent Systems, 70569 Stuttgart, Germany
[2]Laboratory for Soft Bioelectronic Interfaces, École Polytechnique Fédérale de Lausanne, 1202 Geneva, Switzerland
[3]Faculty of Medicine and Health Technology, Tampere University, 33720 Tampere, Finland

[†]Correspondence to: sitti@is.mpg.de
*Equally contributing first authors

**I** 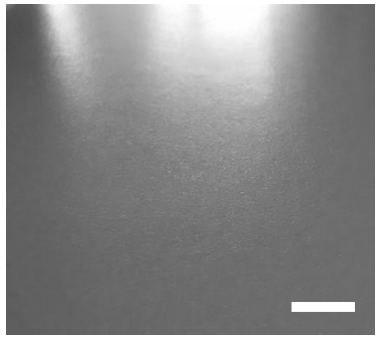

**II** 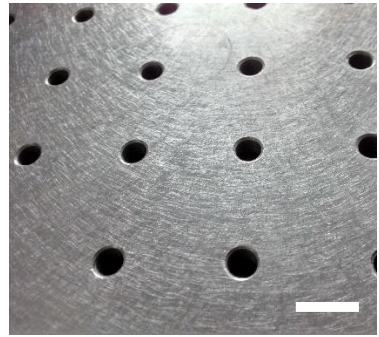

**III** 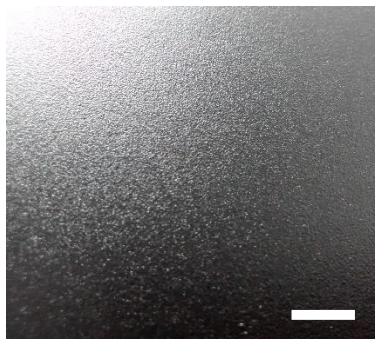

**IV** 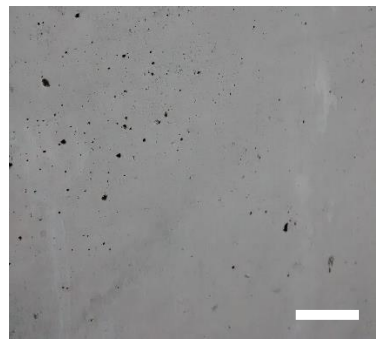

**Fig. S1 | Top-view photographs of the rough surfaces used in the attachment tests.** (**I**): RS 1 ceramic, (**II**): RS 2 steel, (**III**): RS 3 polymer, and (**IV**): RS 4 concrete. Scales indicate 10 mm.

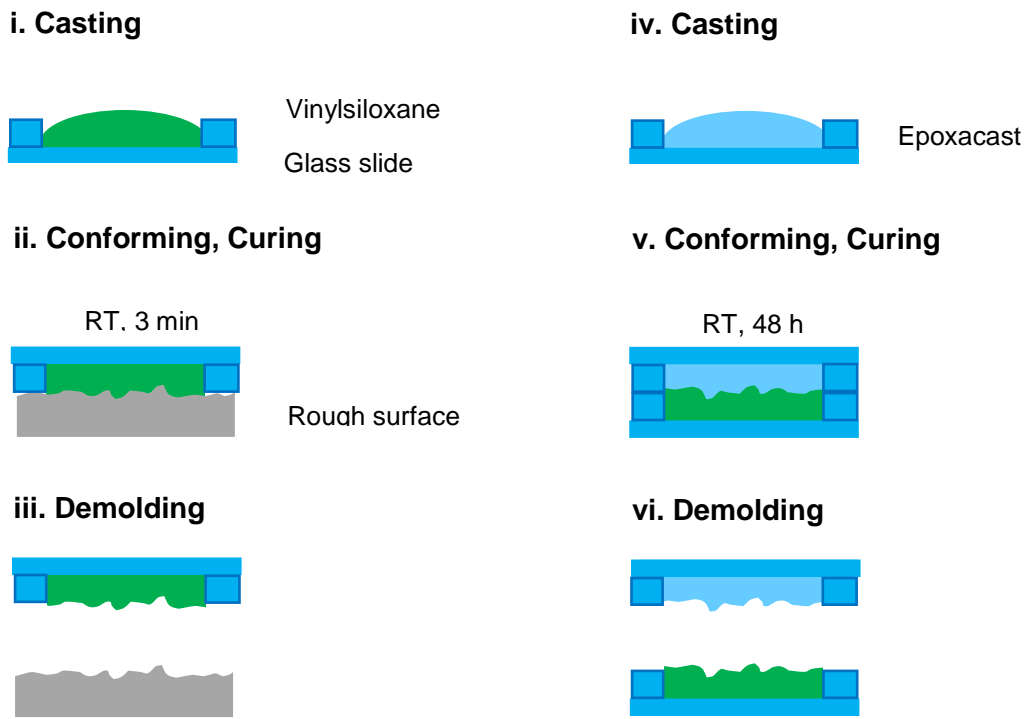

**Fig. S2 | Fabrication process of the rough surface replicas.** Schematic demonstrates the individual fabrication steps. Fabrication of the vinylsiloxane negative mold (steps i–iii), molding and demolding of the positive rough surface replica made of the epoxy polymer (EpoxAcast™ 690, Smooth-on Inc.) (steps iv-vi).

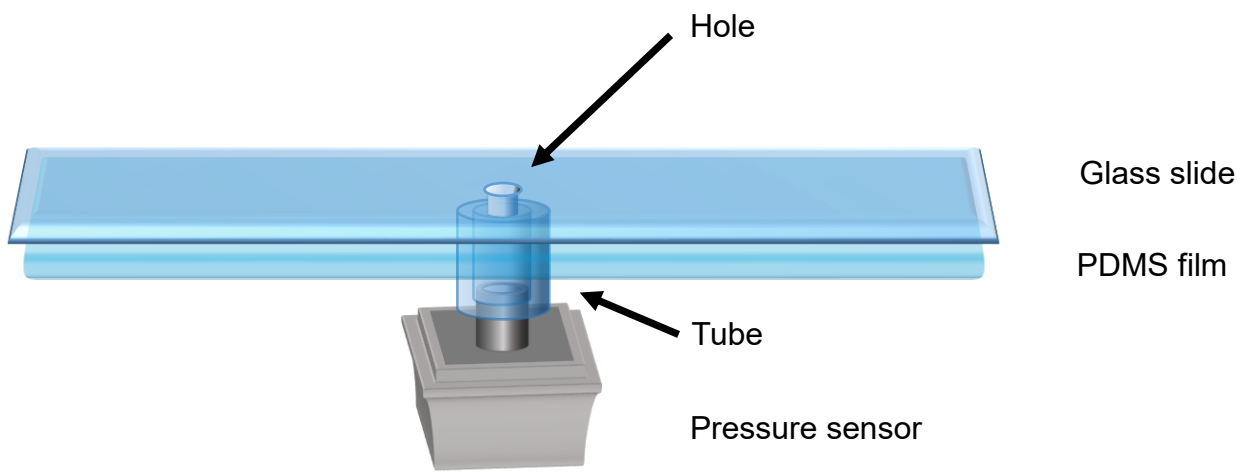

**Fig. S3 | Schematic of the setup for measuring the interfacial pressure.** A pressure sensor is located below the 500 µm diameter hole in the glass slide and attached via a silicon tube. The tube is embedded in a PDMS film, which ensures proper attachment to the glass slide.

**i. 3D Printing**

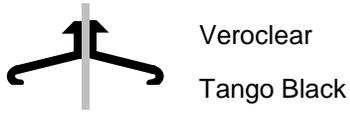

Veroclear
Tango Black

**ii. Casting, Curing at RT**

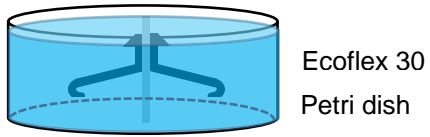

Ecoflex 30
Petri dish

**iii. Demolding**

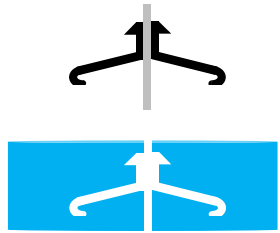

**iv. Silanization**

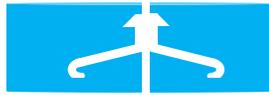

**v. Mold preparation**

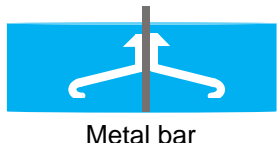

Metal bar

**vi. Injection, Curing at RT**

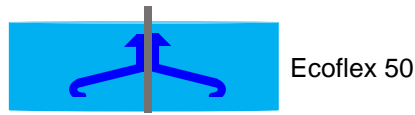

Ecoflex 50

**vii. Demolding**

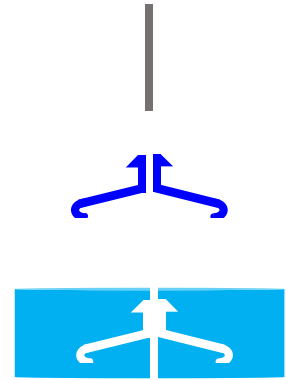

**Fig. S4 | Fabrication process of the soft gripper body.** Schematic demonstrates the individual fabrication steps. Fabrication of the Ecoflex 30 negative mold (steps i–iv), molding and demolding of the positive gripper body made of Ecoflex 50 (steps v-vii).

## A

**i. Casting**

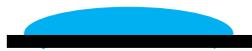

Polymer

Silicon wafer

**ii. Thin film fabrication**

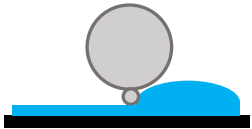

Bar coater

**iii. Curing at RT**

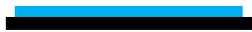

## B

**i. Thin film fabrication**

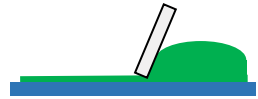

Film applicator

Thin vinylsiloxane film

Glass plate

**ii. Inking**

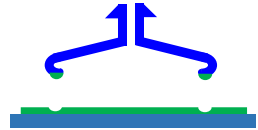

**iii. Bonding, Curing at RT**

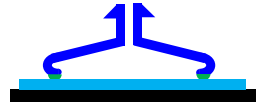

**Fig. S5 | Fabrication process of the gripper membrane and the gripper assembly.** Schematic shows the individual fabrication steps. Fabrication of the membranes **A** (steps i–iii) and bonding of the gripper to the membrane **B** (steps i-iii).

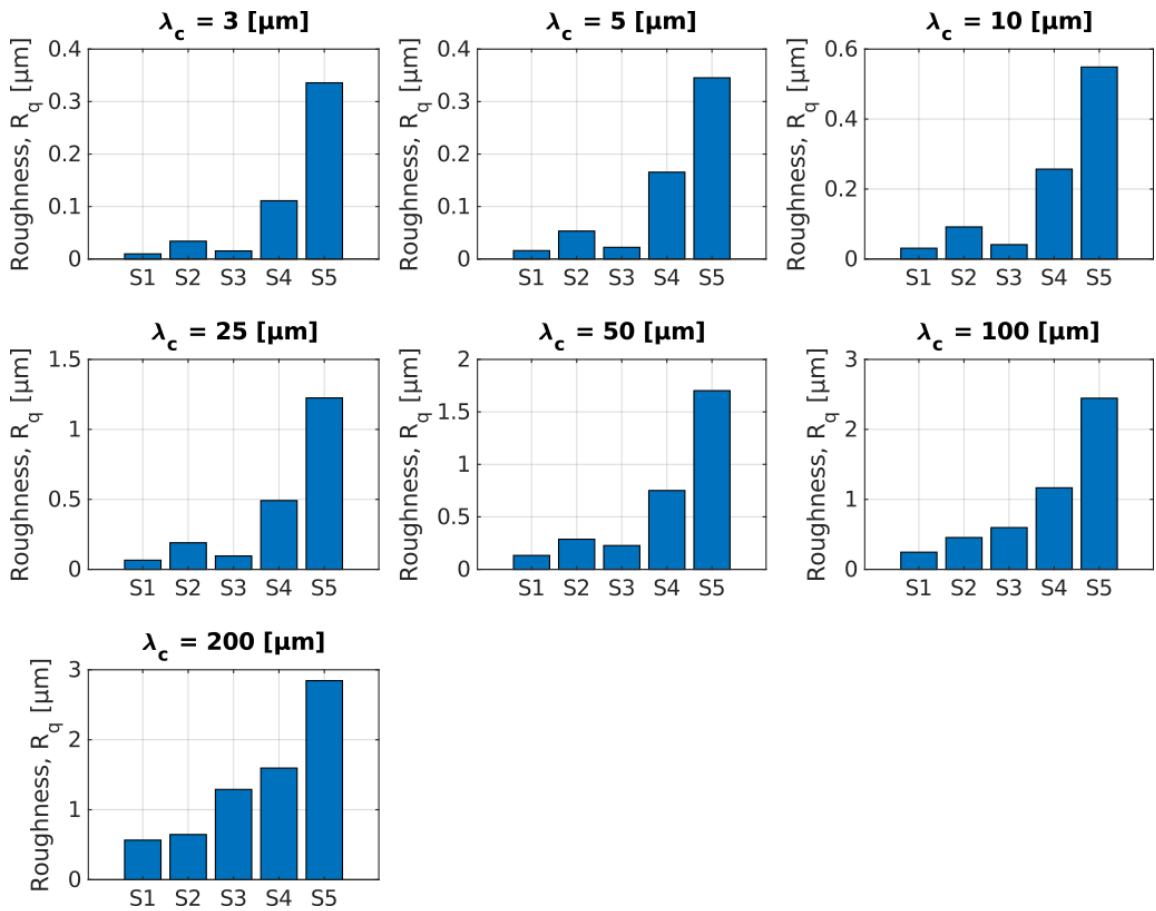

**Fig. S6 | Experimentally measured RMS roughness values of the tested five different roughness replicas (S1-S5) depending on different cut-off wavelengths ($\lambda_c$).** The trend from microscopic (3 μm) to macroscopic (200 μm) RMS roughness values shows that the tested rough surfaces are composed of various length-scale micro- and macroscopic roughness values.

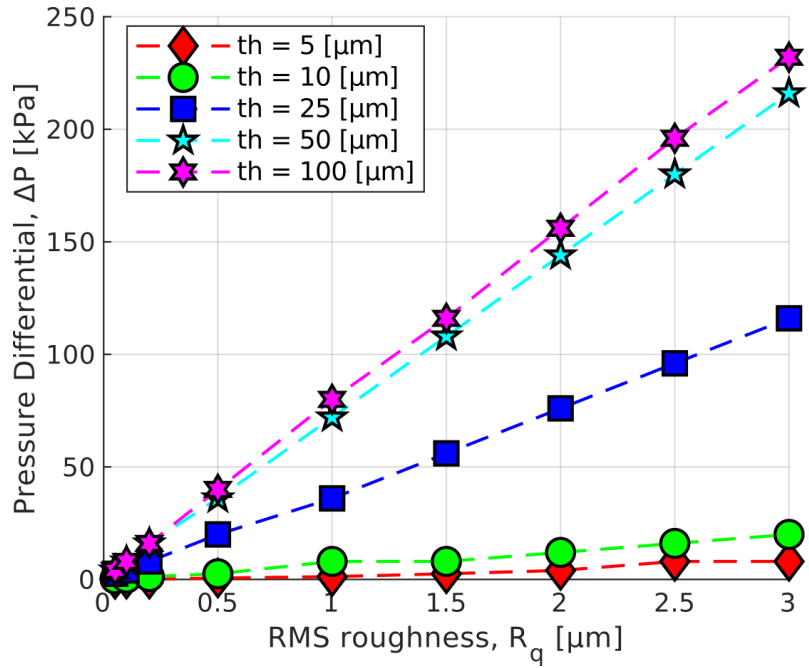

**Fig. S7 | Conformability of the PDMS membrane on a rough surface with the macroscopic rough structures ($\lambda_c = 100$ μm) as a function of the membrane thickness (5-100 μm).** The experimental results suggest that the surface conformability is enhanced by reducing the thickness of the PDMS membrane. This allows the SSG with the PDMS membrane thinner than 10 μm to grip RS 3.